\documentclass[letterpaper,iop,letter]{emulateapj}
\pdfoutput=1
\usepackage{thumbpdf} 
\usepackage{amsmath,amssymb} 
\usepackage{graphicx,mathptmx} 
\usepackage{natbib} 
\usepackage[usenames]{color} 
\usepackage{ifpdf}  
\include{random} 

\usepackage{xcolor}[2007/01/21] 
\usepackage{refcount} 
\usepackage{textcase} 
\usepackage{epstopdf} 

\ifpdf
\pdfoutput=1 
\else

\fi

\newcommand{\beq}{
\begin{equation}
}
\newcommand{\eeq}{
\end{equation}
}
\newcommand{\beqa}{
\begin{eqnarray}
}
\newcommand{\eeqa}{
\end{eqnarray}
}

\newcommand{\kgfigbeg}[1]{
\begin{figure}
\hypertarget{#1}{}%
}
\newcommand{\kgfigend}[2]{
\label{f:#1}
\end{figure}
\bookmarksetup{color=[rgb]{0.54,0,0}}
\bookmark[rellevel=1,keeplevel,dest=#1]{Fig \ref*{f:#1}: {#2}}
\bookmarksetup{color=black} 
}

\newcommand{\kgfigstarbeg}[1]{
\begin{figure*}[ht!]
\hypertarget{#1}{}%
}
\newcommand{\kgfigstarend}[2]{
\label{f:#1}
\end{figure*}
\bookmarksetup{color=[rgb]{0.54,0,0}}
\bookmark[rellevel=1,keeplevel,dest=#1]{Fig \ref*{f:#1}: {#2}}
\bookmarksetup{color=black} 
}

\newcommand{\kgtabbeg}[1]{
\begin{deluxetable}{#1}
}
\newcommand{\kgtabend}[2]{
\label{t:#1}
\end{deluxetable}
\bookmarksetup{color=[rgb]{0,0,0.54}}
\bookmark[
rellevel=1,
keeplevel,
dest=table.\getrefnumber{t:#1}
]{Table \ref*{t:#1}: #2}
\bookmarksetup{color=[rgb]{0,0,0}}
}

\newcommand{\kgtabstarbeg}[1]{
\begin{deluxetable*}{#1}
}
\newcommand{\kgtabstarend}[2]{
\label{t:#1}
\end{deluxetable*}
\bookmarksetup{color=[rgb]{0,0,0.54}}
\bookmark[
rellevel=1,
keeplevel,
dest=table.\getrefnumber{t:#1}
]{Table \ref*{t:#1}: #2}
\bookmarksetup{color=[rgb]{0,0,0}}
}

\newcommand{\units}[1]  {\ensuremath{\mathrm{\ {#1}}}}
\newcommand{\msun}     {\ensuremath{{{M}}_{\scriptscriptstyle \odot}}}

\providecommand{\ion}[2]{#1$\;$\textsmaller{\@Roman{#2}}}

\newcount\mydice
\setrannum{\mydice}{1}{10}  

\newcommand{\jonmiller} {
\ifcase\mydice%
        \or  {{\"J}on M.\ Miller}%
        \or  {J{\"o}n M.\ Miller}%
        \or  {Jo{\"n} M.\ Miller}%
        \or  {Jon {\"M}.\ Miller}%
        \or  {Jon M.\ {\"M}iller}%
        \or  {Jon M.\ M{\"i}ller}%
        \or  {Jon M.\ Mi{\"l}ler}%
        \or  {Jon M.\ Mil{\"l}er}%
        \or  {Jon M.\ Mill{\"e}r}%
        \or  {Jon M.\ Mille{\"r}}%
    \fi%
}
\newcommand{\justmiller} {
\ifcase\mydice%
        \or  {Miller}%
        \or  {Miller}%
        \or  {Miller}%
        \or  {Miller}%
        \or  {{\"M}iller}%
        \or  {M{\"i}ller}%
        \or  {Mi{\"l}ler}%
        \or  {Mil{\"l}er}%
        \or  {Mill{\"e}r}%
        \or  {Mille{\"r}}%
    \fi%
}

\def\spose#1{\hbox to 0pt{#1\hss}}
\newcommand{\lta}{\mathrel{\spose{\lower 3pt\hbox{$\mathchar"218$}}
      \raise 2.0pt\hbox{$\mathchar"13C$}}}
\newcommand{\gta}{\mathrel{\spose{\lower 3pt\hbox{$\mathchar"218$}}
      \raise 2.0pt\hbox{$\mathchar"13E$}}}
\def\simlt{\mathrel{\rlap{\lower 3pt\hbox{$\sim$}}\raise 2.0pt\hbox{$<$}}}
\def\simgt{\mathrel{\rlap{\lower 3pt\hbox{$\sim$}} \raise 2.0pt\hbox{$>$}}}

\makeatletter
\let\jnl@style=\rmfamily
\def\ref@jnl#1{{\jnl@style#1}}%
\newcommand\icarus{\ref@jnl{Icarus}}%
\makeatother

\definecolor{KayhanCiteColor}{rgb}{0,0.08,0.35}
\definecolor{KayhanURLColor}{rgb}{0,0.08,0.35}
\definecolor{KayhanLinkColor}{rgb}{0,0.08,0.35}
\definecolor{KayhanPageColor}{rgb}{0,0.08,0.35}
\definecolor{medred}{rgb}{0.75,0.0,0.0}

\shorttitle{Observing Gaps and Holes in Black Hole Accretion Disks}
\shortauthors{G\"{u}ltekin \& \justmiller}

\submitted{Accepted by The Astrophysical Journal}

\makeatletter
\ifx\undefined\hyperref
\usepackage[pdftitle={Observational consequences of merger-driven gaps in black hole accretion disks},pdfauthor={Kayhan Gultekin},pdfsubject={Astrophysics},pdfkeywords={black hole physics --- galaxies: general --- galaxies: nuclei --- galaxies: bulges --- methods: statistical},letterpaper,linktocpage,colorlinks=true,linkcolor={KayhanLinkColor},urlcolor={KayhanURLColor},citecolor={KayhanCiteColor},hyperfootnotes=false]{hyperref}
\else\relax\fi
\makeatother
\usepackage[figure,figure*]{hypcap} 
\usepackage{atveryend} 
\usepackage[
  open,
  openlevel=3,
  atend
]{bookmark}[2010/04/08]  
\begin{document}

\label{firstpage}
 
\title{Observable Consequences of Merger-Driven Gaps and Holes in Black Hole Accretion
Disks}

\author{Kayhan G\"{u}ltekin\altaffilmark{1}}
%
\author{\jonmiller\altaffilmark{1}}
\affil{\altaffilmark{1}Department of Astronomy, University of Michigan, 500 Church Street, Ann Arbor, MI 48109, USA
\href{mailto:kayhan@umich.edu}{kayhan@umich.edu}, \href{mailto:jonmm@umich.edu}{jonmm@umich.edu}.}

\begin{abstract}
We calculate the observable signature of a black hole accretion disk
with a gap or hole created by a secondary black hole embedded in the
disk.  We find that for an interesting range of parameters of black
hole masses ($\sim10^6$--$10^9\ \msun$), orbital separation ($\sim1\
\units{AU}$ to $\sim 0.1\ \units{pc}$), and gap width (10--190 disk
scale heights), the missing thermal emission from a gap manifests
itself in an observable decrement in the spectral energy distribution.
We present observational diagnostics in terms of power-law forms that
can be fit to line-free regions in AGN spectra or in fluxes from
sequences of broad filters.  Most interestingly, the change in slope
in the broken power-law is almost entirely dependent on the width of
gap in the accretion disk, which in turn is uniquely determined by
mass ratio of the black holes, such that it scales roughly as
$q^{5/12}$.  Thus one can use spectral observations of the continuum
of bright active galactic nuclei to infer not only the presence of a
closely separated black hole binary but also the mass ratio.  When the
black hole merger opens an entire hole (or cavity) in the inner disk,
the broad band SED of the AGN or quasar may serve as a diagnostic.
Such sources should be especially luminous in optical bands but
intrinsically faint in X-rays (i.e., not merely obscured).  We briefly
note that viable candidates may have already been identified, though
extant detailed modeling of those with high quality data have not yet
revealed an inner cavity.

\end{abstract}

\section{Introduction}
\label{intro}

The prevalence of supermassive black holes (SMBHs) at the centers of
galaxies is well established \citep{richstoneetal98}.  In the context
of a hierarchical merging universe, merging galaxies can lead to the
merging of the black holes \citep[BHs;][]{vhm03}.  Mergers of BHs are
important as one potential pathway for the growth of BHs and have even
been suggested as the principal cause of the scaling relations between
BH mass and host galaxy properties \citep{2011ApJ...734...92J}.
Mergers are also strong gravitational wave emitters, and the
asymmetric emission of gravitational waves, especially from spinning
BHs, can lead to a recoil of the merged BH large enough to kick it out
of the host galaxy potentially influencing the BH occupation fraction
of galaxies \citep[e.g.,][]{2008MNRAS.384.1387V, 2008ApJ...682L..29B,
merrittetal04}.  Even if two galaxies with BHs merge and the BHs sink
to the center of the merged galaxy, however, it is not a given that
the BHs will merge within a Hubble time
\citep[e.g.,][]{2003ApJ...596..860M, 1980Natur.287..307B}.

  The search for precursors to BH mergers has naturally focused on
active galactic nuclei (AGNs) that are either spatially resolvable at
separations of $\sim1\,\units{kpc}$
\citep[e.g.,][]{2012ApJ...753...42C, 2012arXiv1201.1904B} or have
spectroscopically distinct broad line regions at separations of
$\sim0.1\,\units{pc}$ \citep[e.g.,][]{1996ApJ...464L.107G, 2009Natur.458...53B,
2010ApJ...716..866S, 2011arXiv1106.2952E}.

One avenue for finding BH pairs at very small separations comes from
an analogy with protoplanetary disks in which the presence of a planet
may be inferred from its influence on the protoplanetary disk,
particularly through gaps and holes in the disk carved out by the
planet \citep{1980ApJ...241..425G, 1996ApJ...460..832T,
2005astro.ph..7492A, 2008ApJ...682L.125E}.  Here, we
consider the observable consequences of an accretion disk gap caused
by a secondary BH as has been considered before
\citep{1991MNRAS.250..505S, 2009ApJ...700.1952H, 2010MNRAS.407.2007C}.
Owing to the complexity of AGN and quasar spectra, we have developed
observational diagnostics in terms of power-law forms that can be fit
to line-free regions in spectra, or potentially even to fluxes derived
from sequences of broad filters.  We have also given consideration to
signatures that may be evident in broad-band SEDs of quasars and AGN
that span the optical and X-ray regimes.

In section \ref{setup} we describe our assumptions
regarding the accretion disk and secondary BH in which we are
interested.  Then we outline the conditions for a gap to be opened and
its subsequent evolution in section \ref{gapopening} and
\ref{gapevolution}, respectively.  In section \ref{gapsed} we present
our calculations of the spectral energy distribution from a gapped
accretion disk and the physical inferences that can be made from
observations.  In section \ref{holesed} we describe the basic
electromagnetic appearance of an accretion disk with a large inner
hole and suggest some candidate sources.  Finally, we discuss our
results and future work in section \ref{discuss}.

\section{Description of the physical system}
\label{setup}

The presence of a (proto-)planet in a
proto-planetary disk  in orbit around a central star can lead to
transfer of angular momentum such that the planet moves inwards \citep[Type I migration;][]{1980ApJ...241..425G}.  Under certain
circumstances (section \ref{gapopening}), the planet can clear
material near its radius, opening up a gap.  Dynamical interactions
between the planet and material at the gap edges tend to drive both the gap
and the planet inward 
\citep[type II migration;][]{1996ApJ...460..832T}.

We are interested in the observational consequences of analogous
systems for SMBHs.  Consider a central BH of
mass $M$ with an accretion disk around it with a secondary BH
of mass $m_s$, mass ratio $q \equiv m_s / M$, in the plane of the disk at an
orbital radius of $a$.  If the secondary BH can open a gap in
the accretion disk, the relative difference in emission would create
an observational signature that could be used as evidence of close
SMBH binary pairs at separations of $\sim10^{-3}\,\units{pc}$.

We assume that the accretion disk is as described by
\citet{1973A&A....24..337S}, i.e., an $\alpha$-disk.  The
$\alpha$-disk model is necessarily a mathematically convenient
approximation rather than a fully physical model, but it is still a
successful model in many respects for a wide variety of astrophysical
accretion systems.  Our results are not bound to the
\citet{1973A&A....24..337S} disk description, and we consider
variations in section \ref{gapsed}.  What we require is that (i)
azimuthally averaged disk midplane temperature is monotonic with
radius in the inner $\sim10^3R_G$, where $R_G \equiv GMc^{-2}$ and (ii)
$h/r$, the ratio of disk scale height to radius, is roughly constant.
The recent considerations of azimuthal variations in disk temperature
\citep{2011ApJ...727L..24D, 2012arXiv1206.0739D} still assume that the
scale of the variability does not change with radius so that for a
given variability model, each azimuthally averaged annular emission is
still a function only of $r$.  A separate issue is that tidal
interaction between the binary black holes and gas disk can lead to
spiral structures in either the gapped phase
\citep{2005astro.ph..7492A} or the circumbinary phase
\citep[e.g.,][]{2005MNRAS.364L..56R, 2009MNRAS.393.1423C}.  The spiral
structure may be (but also may not be) averaged out in an azimuthal
sense.  For the sake of simplicity, we ignore any potential spiral
structure in this paper.  The secondary hole is assumed to have
already found its way into the accretion disk of the primary and
settled into a coplanar orbit as has been shown to occur
\citep{2007ApJ...661L.147B, 2010MNRAS.402..682D}.
As we discuss below, the mass ratios of interest are $q>0.003$.  Below
about $q=0.1$--$0.01$, a potential concern is that the mergers with
halos containing small black holes will lead to a complete dispersal of
the secondary halo before the BH can settle to the center.  Recent
numerical simulations, however show that it is possible for mergers
with dwarf satellites can lead to secondary BHs of mass ratios
$q=0.01$--$0.001$  getting close enough to the
primary that local processes can bring the BHs together
\citep{2010ApJ...721L.148B}.  It is also possible that
intermediate-mass BHs could form in situ in the accretion disk and
would naturally be coplanar and have small mass ratios
\citep{2012arXiv1206.2309M}.

\section{Conditions for gap opening}
\label{gapopening}

There are two requirements that must be met in order for a gap in the
disk to open: (1) the Hill sphere radius of the secondary must be
larger than the disk scale height and (2) gap closing timescale from
the viscous reaction of the disk is longer than the gap opening
timescale.  Our calculations in this and the following section are
inspired by and dependent on the very closely analogous calculations
for proto-planetary disks \citep[e.g.,][]{1980ApJ...241..425G,
1986ApJ...309..846L, 1994ApJ...421..651A, 1996ApJ...460..832T,
1996ApJ...467L..77A, 2006Icar..181..587C}.

\subsection{Hill Sphere}
\label{hillcalc}

The Hill sphere defines the region around an orbiting body in which it
is gravitationally dominant and has radius:
\begin{equation}
R_H\equiv{r}\left(\frac{m_s}{3M}\right)^{1/3}=r\left(\frac{q}{3}\right)^{1/3}.
\label{hilldef}
\end{equation}
The secondary BH will clear a gap in the accretion disk if
$R_H\ge{h}$, which happens only when $q{\ge}3(h/r)^3$.  In the gas
pressure dominated region where the binary orbits are most detectable,
$h/r=0.004$--0.008 for thin disks.  We take a conservative value of
$h/r=0.01$ as our fiducial value implies that $q\ge3\times10^{-6}$ and
even minor mergers are sufficient to carve a gap of width $w\sim2R_H$.

\subsection{Gap closing}
\label{gapclosing}
Gaps will only exist if the gravitational torques open them faster
than viscous diffusion can close them.  The timescale for opening a
gap is 
\begin{equation}
t_\mathrm{open}\sim\frac{1}{\ell^2q^2\Omega}\left(\frac{w}{r}\right)^2,
\end{equation} where $\Omega$ is the angular frequency of the secondary and
$\ell=r{\Omega}c_s^{-1}$ is the order of the Linblad resonance driving
the gap creation \citep{1996ApJ...460..832T}.  The timescale to close
the gap is
\begin{equation}
t_\mathrm{close}\sim\frac{w^2}{\nu},
\end{equation}
where $\nu$ is the kinematic viscosity.  
Requiring $t_\mathrm{close}>t_\mathrm{open}$, yields the requirement that
\begin{equation}
q>\left(\frac{c_s}{r\Omega}\right)^2\alpha^{1/2}=\left(\frac{h}{r}\right)^2\alpha^{1/2},
\end{equation}
where $\Omega$, $c_s$, and $h$ are all taken to be at the radius of
the secondary.  This requirement is equivalent to $q>10^{-5}$ for our
fiducial disk parameters.  Alternative estimates to the critical mass
ratio are possible \citep[e.g.,][]{1986ApJ...309..846L, 1997Icar..126..261W,
2006Icar..181..587C, 2012arXiv1205.4714K}.
For example, one may estimate the gap width by balancing angular
momentum transfer rates via gravitational torques by the secondary on
the disk with that via viscous diffusion from the disk.  Then,
requiring the gap width to be larger than the disk scale height or
larger than the Hill sphere leads to estimates such as $q > A^{1/2}
(h/r)^{5/2} \alpha^{1/2}$ or $q > (A/3) (h/r)^2 \alpha$, respectively,
where $A \approx 50$ is a factor relating to the geometry and the sum
of the torque modes derived from numerical simulations.  Such
calculations give comparable estimates.

\subsection{Longevity of the gap}
\label{longevity}

In order for a gapped accretion disk to be of any observational
consequence, it needs to persist for a reasonably large fraction of an
AGN's lifetime.  If the angular momentum transferred from the
secondary to the outer disk is greater than the angular momentum
transferred from the inner disk to the secondary, then the secondary
and the gap will both migrate inwards.  The gap migration timescale when
the local disk mass dominates over the secondary mass is
\citep{1996ApJ...460..832T}
\begin{equation}
t_\mathrm{migr}=\frac{2}{3\alpha}\left(\frac{h}{r}\right)^{-2}\Omega^{-1}.
\end{equation}
For a $10^8\,\msun$ primary with a secondary at $\sim10^3R_G$, this
corresponds to $\sim3\times10^5\,\units{yr}$.  For secondaries with
masses higher than our fiducial value such that the secondary mass
dominates over the local disk mass, the gap migration time is
lengthened to $\sim10^6\,\units{yr}$, very similar to the timescale
calculated by \citet{2009ApJ...700.1952H}.  Our timescale calculation
neglects the fact that gas may accumulate at the outer edge of the
gap, causing the gap width to decrease and potentially even close
\citep{2012arXiv1205.4714K, 2012arXiv1205.5268K}.  If the gap does not
close all the way, the lifetime would be extended by a factor of $\sim
10$, though at reduced observability due to the reduced width for
fixed mass ratio.

When the BHs are close to each other, gravitational radiation is
strong enough to shrink the orbit on interesting timescales.  The
\citet{peters64} orbit-averaged evolution of the semimajor axis of two
masses in orbit around each other with semimjaor axis $a$ and
eccentricity $e$ is
\begin{equation}
\frac{da}{dt}=-\frac{64G^3Mm_s(M+m_s)}{5c^5a^3(1-e^2)^{7/2}}\left(1+\frac{73}{24}e^2+\frac{37}{96}e^4\right),
\end{equation}
where $G$ is the gravitational constant and $c$ is the speed of light.
For the case we consider, $e=0$.  The width of the gap is $R_H$, and
the timescale for the secondary BH to move out of the gap is
%
\begin{equation}
T_g=R_H\left(\frac{da}{dt}\right)^{-1}=\frac{5c^5a^4}{64G^3Mm_s(M+m_s)}\left(\frac{m_s}{3M}\right)^{1/3}.
%
\end{equation}
Taking $a=r=10^3R_G$ and $q=10^{-2}$, then
$T_g\approx2\times10^{7}(M/10^8\,\msun)\,\units{yr}$, much longer than
other relevant timescales.

In order to predict the actual rate of observable gapped disks, the
actual rate of tight binary formation and their incidence with respect
to the start of AGN activity are needed, but these are unknown.  We
can, however, predict the maximum fraction of an AGN's duration that a
binary in an observable configuration can last, assuming that the
binary companion of the right mass reaches an observable separation.
The lifetime of an AGN is uncertain, but a typical value may be
$\sim10^7\,\units{yr}$ \citep{2004cbhg.symp..169M}.  For a gap
lifetime of $\sim10^6\,\units{yr}$, then up to $10^{-1}$ of a binary
AGN's duty cycle can be in an observable gapped state.  If binary
acquisition and AGN activity are anti-correlated, then the implied
observability would be much lower.

\section{Evolution of the gap}
\label{gapevolution}

Once a gap is established it will evolve along three possible routes
when considering only viscous effects.  At the beginning it is a
gapped disk, effectively a standard disk with an annulus of width $w$
missing.  If the gap migration timescale is faster than or comparable
to the timescale for accretion of the disk material inside of the gap,
then the gap will have constant $w/h$ with a decreasing radius.  

If the mass accretion rate for the inner disk is faster than the gap
migration rate, then it will partially decouple from the outer disk.
This will first lead to an effective widening of the gap from the
inner edge, making it more detectable.  The inner disk will be
continuously depleted and replaced by any material that accretes
across the gap.  If the mass accretion rate across the gap is the same
as if there were no gap, then there would be no difference.  On the
other hand, if the accretion rate is a small fraction of what it would
have been otherwise, the inner disk would be replaced by an accretion
disk of a smaller accretion rate.

The details of how much accretion will persist across a gap created by
a satellite are the subject of current numerical simulations
\citep[e.g.,][]{2012MNRAS.420..860S, 2012ApJ...749..118S,
2012ApJ...755...51N}.  For the small mass ratios that are our primary
interest here, however, it is plausible that the accretion rate is
only reduced to 10\% of what it would be in the absence of a
satellite, i.e., $\dot{m}_\mathrm{red}=0.1\dot{m}$, where
$\dot{m}\equiv\dot{M}/\dot{M}_\mathrm{Edd}$ and
$\dot{M}_\mathrm{Edd}\equiv{L}_\mathrm{Edd}\eta^{-1}c^{-2}\approx3\times10^{-8}(M/\msun)\,\msun\,\units{yr^{-1}}$
for an assumed $\eta=0.1$.

For higher mass ratios, it is plausible that very little matter will
accrete past the gap, resulting in essentially no replacement of the
inner gap.  In such a case, instead of a gap, there would be a hole in
the central accretion disk.  The observational consequences of an
accretion disk with a central hole---and the possible electromagnetic
signal once the BHs merge---have been discussed before
\citep{2005ApJ...622L..93M, 2009ApJ...700..859O, 2010ApJ...714..404T},
and we return to this in section \ref{holesed}.

The time to accrete the entire inner disk is
\begin{equation}
t_a=M_\mathrm{in}\dot{M}^{-1},
\end{equation}
where $\dot{M}$ is the mass accretion rate and $M_\mathrm{in}$ is the
mass of the inner accretion disk.  For our fiducial accretion disk, we
take the BH to be accreting at a fraction $\dot{m}=0.01$.
The largest orbit we consider is roughly at the boundary outside of
which free-free opacity dominates over Thomson scattering so that the mass 
inside the gap is roughly
\begin{equation}
M_\mathrm{in}\approx6\times10^{-11}\left(\frac{M}{\msun}\right)^{11/5}\alpha^{-4/5}\dot{m}^{23/15}\ \msun
\end{equation}
corresponding to an accretion timescale of
\begin{equation}
t_a\approx0.002\alpha^{-4/5}\left(\frac{M}{\msun}\right)^{6/5}\dot{m}^{8/15}\ \units{yr}.
\end{equation}
For our fiducial case, $t_a\approx3\times10^7\ \units{yr}$, so that
the inner disk is depleting itself more slowly than the gap's inward
migration.  The above is only valid for $q\ll1$.  For mass ratios near
unity, the dynamical torques will clear a hole inside of the orbit on
a short timescale.

\section{Spectral Energy Distribution of a gapped accretion disk}
\label{gapsed}
The spectral energy distribution (SED) of a thermally emitting disk with
temperature a function only of radius is
\begin{equation}
F_\lambda(\lambda)=\int_{R_\mathrm{in}}^{R_\mathrm{out}}B_\lambda[T(r^\prime)]g(r^\prime)2\pi{r^\prime}dr^\prime,
\label{flambdased}
\end{equation}
where $B_\lambda(T)$ is the Planck function and $g(r^\prime)$ is a
function that describes the emissivity as a function of radius.  We
assume that $g(r^\prime)=1$ (optically thick) where the disk material
is present (i.e., $R_\mathrm{in}<r^\prime<r-w$ and
$r+w<r^\prime<R_\mathrm{out}$) and $g(r)=0$ elsewhere.  This
assumption of an ``on or off'' emissivity is simplistic, and recent
theoretical work \citep{2012MNRAS.420..705T, 2012MNRAS.420..860S,
2012arXiv1205.4714K} has examined in closer detail how the surface
density changes close the gap.  \citet{2012arXiv1205.4714K} found that
close to the gap, the surface density will decrease outside of the
Hill sphere.  In most cases, the surface density does not drop all the
way to zero, but it generally drops by over an order of magnitude.  In
the region of the disk immediately exterior to the gap, the surface
density will increase to above the value it would have in the absence
of the disk.  There is a similar enhancement in temperature because of
the added pressure.  Both of these effects of enhanced surface density
and temperature are most prominent for high mass ratios ($q > 0.01$)
and small separations ($r < 100 R_G$).  For the smaller mass ratios
and wider separations that we consider here, this effect is small.
Assuming graybody instead of blackbody emission, as we do, would tend
to increase the observational signature
\citep[e.g.,][]{2010ApJ...714..404T}.  Inverse Compton scattering of
the highest energy photons from the thermal spectrum will alter the
SED, but this effect happens primarily at wavelengths shorter than
$\sim1000\ \mathrm{\AA}$ \citep{1992MNRAS.258..189R}, meaning that it
has little to no effect on the observability of gaps that we consider.

\kgfigstarbeg{seds}
%
\includegraphics[width=0.33\textwidth]{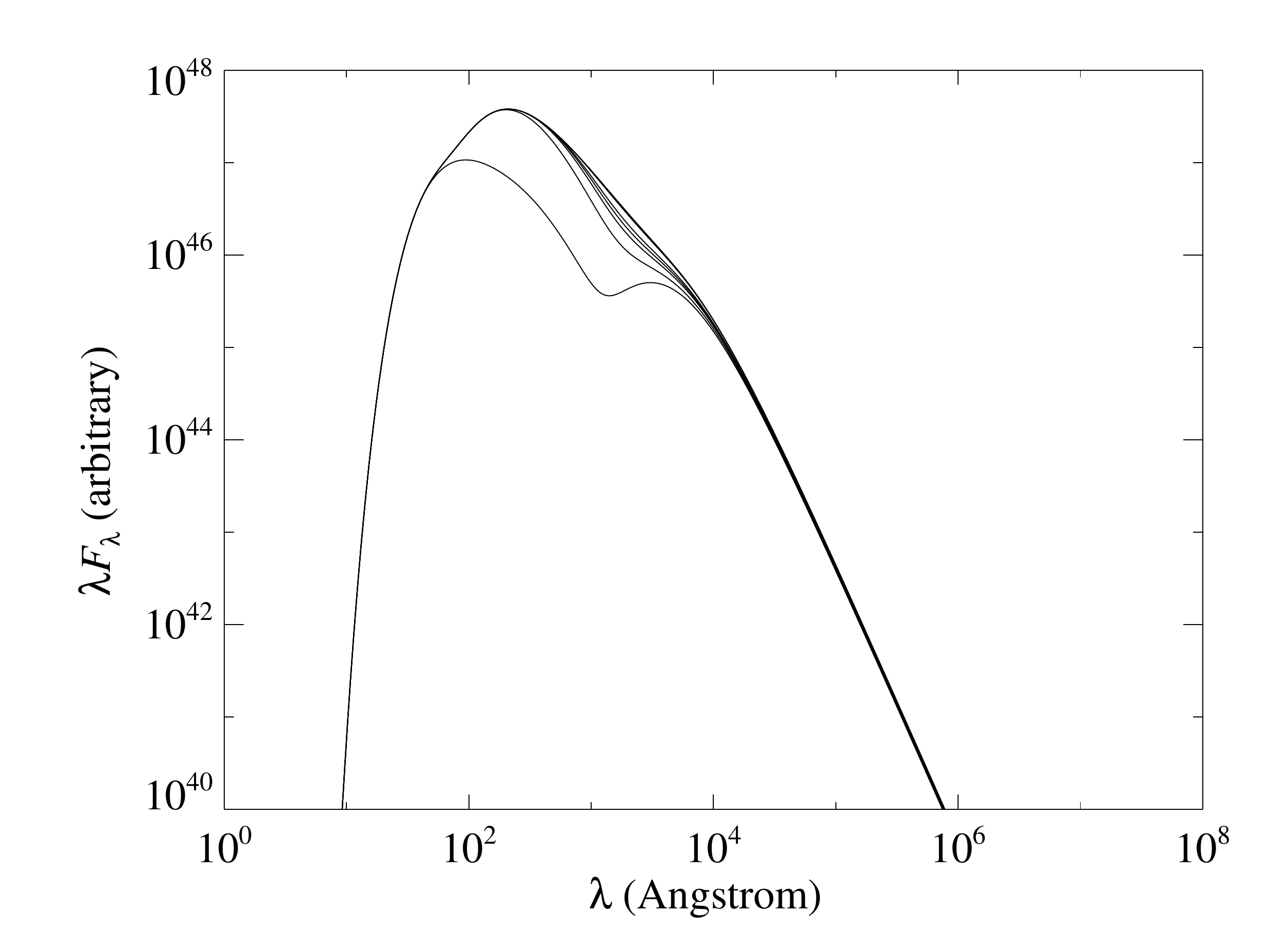}
\includegraphics[width=0.33\textwidth]{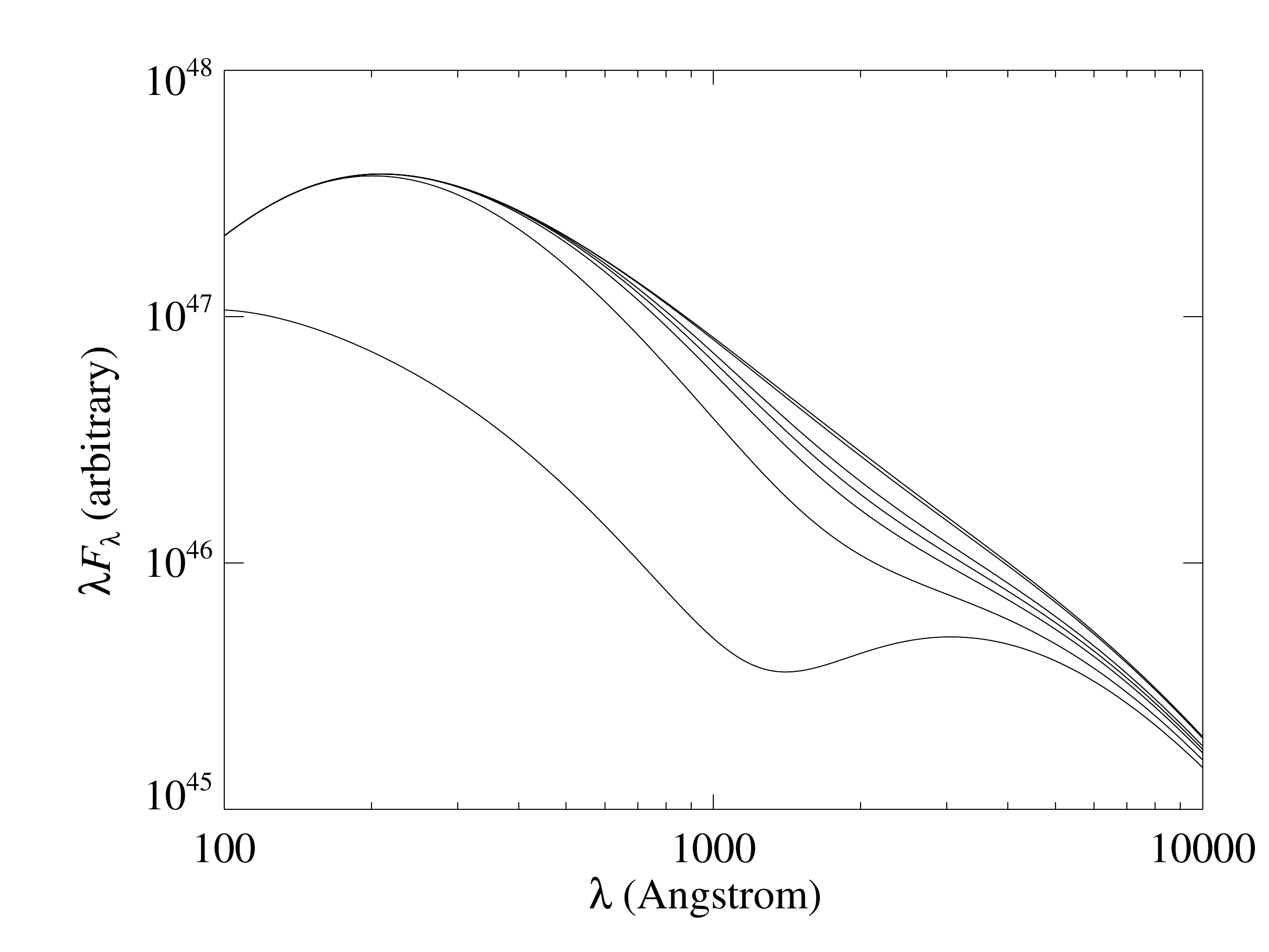}
\includegraphics[width=0.33\textwidth]{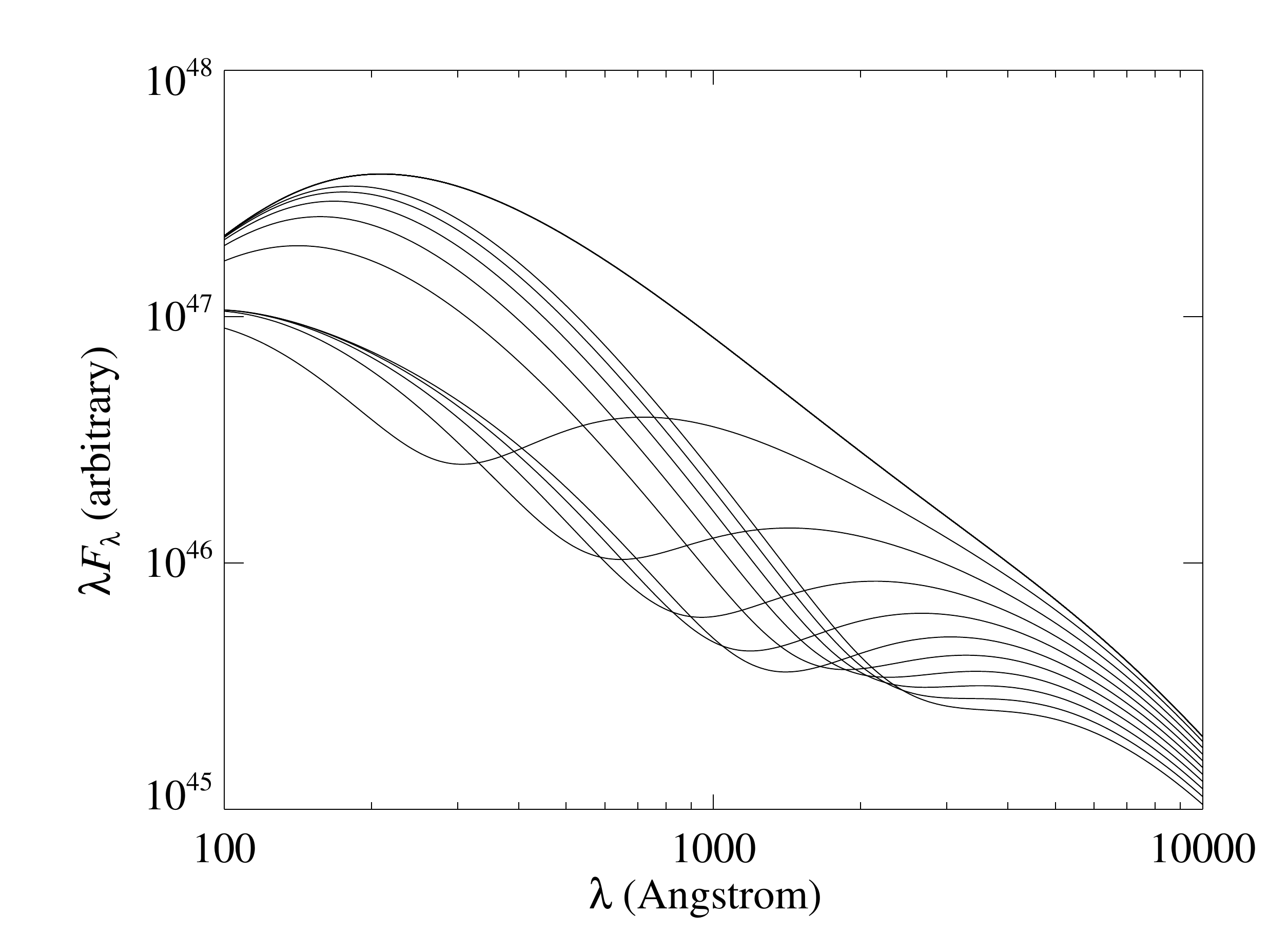}
\caption{All panels show simulated SEDs with gaps, with our fiducial
parameters of $M=10^8\,\msun$, $h/r=0.01$, and assuming our standard
temperature profile.  The left and center panels assume $r=1000R_G$ and
show the effect of increasing the gap width with values $w/h=0$, 10, 60,
80, 100, 140, and 180.  The central panel displays a zoomed in version
of the same and shows the depth of of the dip in the SED increasing as
well as the differences in slopes on either side of the local minimum
increase as $w/h$ increases.  Empirically, we find that the change in
slope scales approximately as $(w/h)^{5/4}$, allowing one to infer the
mass ratio of the binary from just the change in slopes.  The right
panel assumes $w/h=180$ and shows the effect of increasing the gap
radius with values from $r/R_G=200$ to 2000 in steps of 200.  For
reference, also plotted is a gapless SED.  Modeled as a broken
power-law near the gap, we find that the break occurs at a wavelength
that scales with gap radius as $\lambda\sim(r/R_G)^{3/4}$.  This comes
from the temperature profile's dependence of $T\sim{r}^{-3/4}$ and
having the peak wavelength of a blackbody scale as $\lambda\sim
T^{-1}$.  The break wavelength also depends on $M$ and $\dot{m}$ as
shown in equation (\protect{\ref{breaklambda}}).}
\label{seds}
\kgfigstarend{seds}{Simulated SEDs of BH accretion disks with gaps}

The absence of a ring in a thermally emitting disk will show up in the
SED as a broad, shallow dip.  Without a gap, the SED will
fall off as ${\lambda}F_\lambda\sim\lambda^{-4/3}$.  With a gap, the
SED will appear roughly as a broken power-law with a slope that is
steeper and shallower than $\lambda^{-4/3}$ blueward and redward of
the maximum deviation from the standard SED.  While it is obviously
possible to model an observed SED with the proper Planck function
treatment, we use a broken power-law model to demonstrate the precision
required to identify a gapped disk.  We generated SEDs (Fig.\ \ref{seds}) for gapped
disks using equation (\ref{flambdased}) for various values of $w$,
$r/R_G$, $M$, and $\dot{m}$.  For each simulated SED, we fitted broken
power-law in the vicinity of the break of the form
\begin{equation}
\lambda{F}_\lambda=\begin{cases}\lambda^{-x}&\lambda\le\lambda_0\\
\lambda^{-y}&\lambda>\lambda_0\end{cases},
\end{equation}
where $x>4/3$ and $y<4/3$ with continuity and an arbitrary normalization
implied.  The change in slope, $\Delta\equiv{x}-y$, from one
side of the break to the other at $\lambda=\lambda_0$ increases as
the gap width increases.  We find
\begin{equation}
\lambda_0 \approx 140 \left(\frac{r}{R_G}\right)^{3/4} \left(\frac{M}{10^8\,\msun}\right)^{1/4} \left(\frac{\dot{m}}{0.01}\right)^{-1/4} f(w/h)\,\mathrm{\AA},
\label{breaklambda}
\end{equation}
where $f(x)$ is a residual function that is smoothly varying over
$x$ and has absolute value $|f(w/h) - 1| < 0.35$ for most of the
parameter space of interest.
The dependence on $r$, $\dot{m}$, and $M$ is easily understood since
the peak wavelength of a blackbody scales as $\lambda\sim{T}^{-1}$ and
temperature and mass scale as $T\sim\dot{m}^{1/4}r^{-3/4}$ in the
outer regions where the secondary spends most of the time and
$M\sim{T}^{-1/4}$.  The dependence on $w$ comes from the area
weighting of the output SED.  The change in slope is well approximated by
 \begin{equation} 
\Delta\approx0.002(w/h)^{5/4}
\end{equation}
for $w/h<160$.  Above $w/h>160$, the change in slope is larger than
the above approximation because the emitting annulus just outside of
the gap is further down the Wien exponential drop off.  Since $w/h
\sim q^{1/3}$ is purely a function of $q$ under our assumptions, it is
possible to infer the mass ratio of the binary from just the change in
slope, independent of $\dot{m}$, $M$, and $r$.

\citet{1995MNRAS.277..758S} and \citet{2012arXiv1205.5017R} found that
a gapped accretion disk would have an SED such that
$\lambda{F}_\lambda\propto\lambda^{12/7}$.  If $T(r)\sim{r}^{-0.9}$,
then the peak of the SED scales as
$\lambda{F}_\lambda\propto\lambda^{-1.7}$, approximately as
$\lambda^{12/7}$.  There is no qualitative difference in the approach
to modeling SEDs of such gapped disks.  We find that the change in
power-law slope for such a temperature power-law is only decreased by
about 10\% compared to our fiducial temperature profile.  Thus one
could use the overall slope of the continuum power-law to identify
gapped accretion disks and then use detailed measurements of $\Delta$
and $\lambda_0$ to infer properties of the BHs.

\section{SED of an accretion disk with a hole}
\label{holesed}

For higher mass ratios and suitably small separations, almost the
entire region inside of the orbit will be dynamically unstable as can
be seen from considering the size of the Roche lobes for $q\approx1$
binaries.  Thus all gas will be cleared out of this region, and if the
gas cannot be replaced, it will leave an accretion disk with a hole of
size $r\sim2a$ in the center.  Like accretion across the gap, the
ability or inability of the disk to accrete a significant amount of
matter into the hole and the timescale on which it does so is a matter
of debate \citep[e.g.,][]{1994ApJ...421..651A, 1996ApJ...467L..77A,
2005ApJ...622L..93M, 2007PASJ...59..427H, 2009MNRAS.393.1423C,
2010ApJ...714..404T, 2010PhRvD..81b4019S, 2012MNRAS.420..860S,
2012ApJ...749..118S, 2012ApJ...755...51N}.  Because of resonant
interactions between the binary and the gas disk, the orbit of the
secondary can become eccentric---particularly when starting with a
substantial eccentricity---so that accretion onto the secondary is
non-negligible \citep{2011MNRAS.415.3033R}.  Assuming that the inner
hole is mostly cleared of continuum-emitting material and neglecting
the uncertain amount of accretion onto the secondary, the effect of an
inner hole on the SED is much more dramatic than that of a gap.  Most
or all of the shortest wavelength emission will be cut out, leaving a
red SED.

The absence of an inner accretion disk owing to a BH merger event may
therefore be identified through the broad-band SED.  The outer
accretion disk will still have a high accretion rate, and should be
luminous in optical bands.  Indeed, the outer disk may be abnormally
luminous in optical, or even in UV bands \citep[see,
e.g.,][]{2012arXiv1205.5268K}, depending on the mass accretion rate
and the truncation radius.  On simple grounds, UV luminosity is not a
good diagnostic: a holed disk around a BH of lower mass is
observationally degenerate with a filled inner disk around a BH of
higher mass.  Rather, diminished X-ray emission is a better
diagnostic, since non-thermal X-ray emission is thought to tap into
the deep gravitational potential close to the BH
\citep{2012MNRAS.420..705T}.  In supermassive BHs, most X-ray emission
arises from inverse Compton scattering of lower energy photons off hot
electrons in a corona of which relatively little is known.  It is
possible that an X-ray faint AGN could be caused by the absence or
variability of the hot corona \citep{2012arXiv1207.0694M}, but this
requires UV observations to show that there are thermal disk photons
in the UV available to be inverse Compton scattered
\citep[e.g.,][]{2007ApJ...663..103L}.  The amount of X-ray diminishing,
however, depends critically on how much accretion through the hole is
diminished, which has been found in some studies to be only very
mildly altered \citep{2012MNRAS.420..860S}.  If accretion is greatly
diminished, residual accretion within the hole may persist at a low
level, but even this gives rise to X-ray variability in Sgr A*.  Thus,
X-ray variability may be an additional hallmark, which has been
investigated in detail from the perspective of surveys such as the
extended Roentgen Survey with an Imaging Telescope Array
\citep[eROSITA;][]{2012MNRAS.420..860S}.  Thus, a source that is
luminous in optical bands but intrinsically faint (not faint due to
obscuration) and potentially variable in X-rays might signal a recent
or pending BH merger event, including as potential electromagnetic
counterparts to gravitational wave detections from pulsar timing array
experiments \citep{2012MNRAS.420..705T}.

It is possible that such sources have already been identified.  When
the distance to a given AGN is known a priori, e.g., through
spectroscopy, SDSS colors are able to distinguish whether or not a
source is intrinsically luminous in, e.g., red bands, or just
apparently red through obscuration.  In particular, $g-i$ is very
effective, or $\Delta(g-i)$, where one measures color relative to the
mean quasar color in a given red-shift bin.
\citet{2006AJ....132.1977H} employed this technique to 3814 SDSS
quasars with the aim of isolating obscured BHs.  Follow-up
observations with {\it Chandra} revealed that a subset of {\it
intrinsically} red quasars are {\it not} X-ray-faint because of
obscuration.  In particular, seven sources show no evidence of even
moderate intrinsic absorption in X-rays, and two are constrained to
have internal column densities below $2\times10^{21}\,\units{cm^{-2}}$
\citep{2006AJ....132.1977H}.  This suggests that such sources are rare
($7/3814=1.8\times10^{-3}$), consistent with expectations based on the
peculiar configuration required and the relevant timescales, and also
consistent with expectations based on broader surveys
\citep{2008ApJ...685..773G}.  

An interesting example of a quasar that is optically very luminous but
X-ray faint is PHL 1811, but because of the high quality UV data of
this source \citep{2007ApJS..173....1L} and restframe UV observations
of high-redshift analogs \citep{2011ApJ...736...28W,
2011ApJ...743..163L}, detailed models of the SED show it to have a
thermally emitting disk in towards $\sim 6 R_G$, thus lacking an inner
hole \citep{2007ApJ...663..103L}.  This example illustrates that X-ray
faintness is not a \emph{sufficient} condition to infer an accretion
disk hole, but it is probably a \emph{necessary} condition.  The most
promising method for finding holey accretion disks is to select on
X-ray-faint quasars for full SED or spectral modeling in order to
establish the amount of UV and X-ray absorption as well as a rough
estimate of the size of the inner edge of the accretion disk.  For
example, SDSS J094533.99+100950.1 has only upper limits on its X-ray
flux \citep{2011ApJ...736...28W} and very faint UV continuum emission
\citep{2011MNRAS.415.2942C}, but detailed modeling of the broad band
SED and UV spectrum revealed that it is, depending on the extinction
model assumed, heavily extincted with an accretion disk extending into
at least $\sim 10 GMc^{-2}$ \citep{2011MNRAS.415.2942C}.

\section{Discussion and conclusions}
\label{discuss}
We have described the basic observational appearance of a BH
with a gapped or holed accretion disk, both of which are observable
through their broad-band SEDs.  We have focused our consideration on
parameter space that results in observable signatures between 2000\
\AA\ and 2\ ${\mu}m$.  At shorter wavelengths, absorption of
ultraviolet photons will prevent unambiguous determination of details
of the AGN continuum.  At longer wavelengths, reprocessed emission
from dust grains will similarly contaminate thermal emission from an
accretion disk.  In this range, however, it is possible to measure the
continuum emission to the precision needed to infer the presence of a
gap.  Current and future large surveys such as SDSS, Pan-Starrs, and 
LSST may be able to exploit the observational signatures we have 
developed.

The primary outcome of this paper is to delineate clear diagnostics of
accretion disks with gaps and holes.  To do so, we have made some
simplifying assumptions that allow for the clearest picture of what
will happen to the accretion disk SED, but there are, of course, some
potential complications that will need to be considered in the future.
One such complication is that we have treated the emission as only
coming from the top of the accretion disk.  Emission coming from the
walls of the gap will complicate the signal and, depending on the
vertical temperature structure, either increase or decrease the
observability of the gap, potentially by a large amount given the
possibility of dramatically increased $h/r$ at the outer boundary from
gas pile up \citep{2012arXiv1205.4714K}.  Tidal features produced by
the secondary as seen in simulations of protoplanetary disks will also
need to be considered and may lead to additional corroborating
observational signals.  The periods of the secondaries that are
observable are months to several years, allowing for a potential
detection of a variable or modulating signal
\citep{2010AJ....140..642T, 2010ApJ...714..404T, 2012MNRAS.420..860S,
2012ApJ...755...51N}.  One such periodic signature from an accreting
secondary could be \ion{Fe}{25}\ and \ion{Fe}{26} narrow emission
lines that would move with the secondary.  Such a measurement should
be possible with the \emph{Astro-H} X-ray observatory
\citep{2010SPIE.7732E..27T}.

Another potential complication is that the population of observed
quasar SEDs have a large variance \citep[e.g.,][]{2006ApJS..166..470R,
2011ApJS..196....2S} that arises from a number of unknown causes not
thought to be connected to the occurrence of any secondary black hole.
For example, typical optical-to-X-ray spectral indices ($F_\nu \propto
\nu^{\alpha_\mathrm{ox}}$) can range from $\alpha = -1.8$ to $-1.2$.
Unambiguously discerning between low X-ray activity arising from
quasars coming from the low-end of a natural range and low X-ray
activity arising from quasars with a missing inner disk requires
sufficiently deep X-ray observations.  Regarding gapped accretion
disks, our focus on local deviations from an overall power-law
continuum emission mitigates the potential confusion from variance in
SEDs arising from other causes.  That is, it is possible to discern a
dip in a power-law SED portion without \emph{a priori} knowing what
the power-law slope is.  A full theoretical understanding of normal
quasar continuum emission would, of course, allow one to model gaps in
accretion disks with utmost fidelity.

We thank Cole Miller and Mike Eracleous for extremely useful
discussions as well as Zoltan Haiman, Bence Kocsis, Mateusz
Ruszkowski, Alberto Sesana, and Taka Tanaka for thoughtful
comments. We thank Jim Saborio and staff for their hospitality during
which a large portion of this work was completed.  KG acknowledges
support provided by NASA through Chandra Awards GO0-11151X and
G02-13111X and through Hubble Award HST-GO-12557.01-A awarded by the
STScI.

\def\href#1#2{#2}
\bibliographystyle{apjads}
\bibliography{gultekin}
\end{document}